\documentclass[a4paper,fleqn,usenatbib]{mnras}

\usepackage{epstopdf}
\usepackage{wrapfig}
\usepackage{float}
\usepackage{adjustbox}
\usepackage{rotating}
\usepackage{ltablex} 
\usepackage{supertabular,booktabs}
\usepackage{longtable}
\usepackage{graphicx}
\usepackage{amsmath,amssymb}
\usepackage{amsfonts}
\usepackage{float}
\newcommand{\as}{\prime\prime}
\usepackage{multirow}
 \usepackage[normalem]{ulem}
 \useunder{\uline}{\ul}{}
\newcommand{\be}{\begin{equation}} 
\newcommand{\ee}{\end{equation}}
\newcommand{\galex}{{\it GALEX}\,\,}
\newcommand{\astrosat}{{\it AstroSat}\,\,}

\newcommand{\ns}{\!\!}
\def\procspie{{\it Proc. SPIE}\,\,}
\def\apjs{{\it Astrophys.~J.~Suppl.}}

\def\aj{{\it Astronom.~J.\,}}
\def\apjl{{\it Astrophys.~J.~Lett.}}
\def\apss{{\it Astrophys. Space Sci.}}

\def\mnras{{\it Mon.~Not.~R.~Astron.~Soc.} }

\def\aap{{\it Astron.~Astrophys.}\,}

\def\ac{{\it Astron.~Comput.}\,}

\def\pasp{{\it Publ. Astr. Soc. Pac.}\,}
\def\basi{{\it Bulletin of the Astronomical Society of India.}\,}
\title[UVIT Performance]{Investigating the In-Flight Performance of the UVIT Payload on \astrosat}

\author[P.~T.~Rahna et al.]{P.~T.~Rahna,$^{1}$\thanks{E-mail: 7rehanrenzin@gmail.com}
Jayant Murthy,$^{2}$\thanks{E-mail: jmurthy@yahoo.com} 
M.~Safonova,$^{3}$
F.~Sutaria,$^{2}$
S.~B.~Gudennavar$^{1}$
\newauthor
and S. G. Bubbly$^{1}$
\\
$^{1}$Department of Physics and Electronics, CHRIST (Deemed to be University), Bengaluru 560029, India\\
$^{2}$Indian Institute of Astrophysics, Bengaluru 560034, India\\
$^{3}${M.~P.~Birla Institute of Fundamental Research, Bengaluru 560001, India}
}

\begin{document}

\date{Accepted . Received ; in original form }

\pagerange{\pageref{firstpage}--\pageref{lastpage}} \pubyear{2018}

\maketitle

\label{firstpage}

\begin{abstract}
We have studied the performance of the Ultraviolet Imaging Telescope payload on {\astrosat} and derived a calibration of the FUV and NUV instruments on board. We find that the sensitivity of both the FUV and NUV channels is as expected from ground calibrations, with the FUV effective area about 35\% and the NUV effective area about the same as that of \galex\ns. The point spread function of the instrument is on the order of 1.2 -- 1.6\arcsec. We have found that pixel-to-pixel variations in the sensitivity are less than 10\%\ with spacecraft motion compensating for most of the flat-field variations. We derived a distortion correction but recommend that it be applied post-processing as part of an astrometric solution.
\end{abstract}

\begin{keywords}
instrumentation: detectors, techniques: photometric, ultraviolet: general
\end{keywords}
\section{Introduction}

The {\astrosat} satellite was launched by the Indian Space Research Organization (ISRO) on Sept. 28, 2015 into a near-equatorial (inclination 6$^{\circ}$) orbit with an altitude of 650 km \citep{Singh2014}. One of the instruments aboard the satellite is the Ultraviolet Imaging Telescope (UVIT) designed to observe large areas of the sky (field of view: $28'$ diameter) with a resolution better than $1.8^{\as}$ \citep{Kumar2012}.
UVIT was built by the Indian Institute of Astrophysics (IIA) in collaboration with the Inter-University Centre for Astronomy and Astrophysics (IUCAA), the Canadian Space Agency (CSA) and ISRO. It consists of two co-aligned telescopes with three identical intensified CMOS detectors in the  far-ultraviolet (FUV), near-ultraviolet (NUV) and visible (VIS).

The ground calibration of the UVIT has been discussed by \citet{Postma2011}, and the in-flight tests by \citet{Subramaniam2016}. The in-flight calibration of the instrument has been presented by \citet{Tandon2017b}. In this work, we present an independent evaluation of the performance of the UVIT FUV and NUV detectors based on observations taken in the Performance and Verification (PV) phase with additional data from our own observations and later calibration observations. The VIS channel was intended only for tracking purposes and its characteristics will have to be investigated in detail in order to use it for scientific purposes. We will defer this to a further work. The data were reduced and analysed using the {\em JUDE} (Jayant's UVIT Data Explorer: \citet{Murthy_jude2016, murthy2017}) software. Because we have used an independent software system and different calibration techniques, this work provides a verification of the UVIT processing software and the calibration.

\section{Spacecraft and UVIT Instrument}

{\astrosat} was conceived as India's first dedicated astronomy satellite with three X-ray instruments and a UV payload (UVIT) \citep{Pati1998,Pati1999}. A recent overview of the {\astrosat} satellite and its mission was presented by \citet{Singh2014} with a description of the flight configuration of UVIT in \citet{Tandon2017a}. The various payloads on {\astrosat} were turned on in sequence after the launch, with UVIT being the last payload to begin observing on November 30, 2015 with an observation of the open cluster NGC~188.

\begin{figure*}
\includegraphics[scale=0.5]{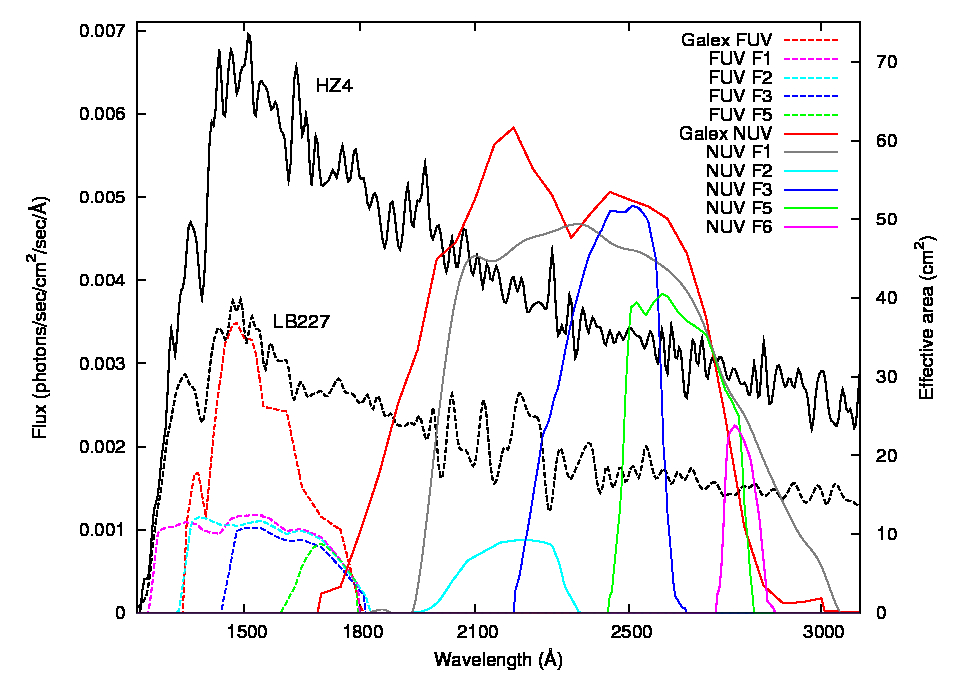}
\caption{Plots of standard stars spectra with over plotted effective areas of UVIT and \galex filters.}
\label{fig:spectra}
\end{figure*}

\begin{table*}
\centering
\caption{General properties of UVIT filters}
\label{filterprop}
\begin{tabular}{cclccc}
\hline 
Filter & Slot (F) & Type & Passband & Effective bandwidth  & Central $\lambda_0$  \\
Channel &  &  & (nm)&  $\Delta\lambda$ (nm) &  (nm) \\     
\hline
\multicolumn{1}{c}{\multirow{6}{*}{FUV}} & 0 & Block    &     &       \\
\multicolumn{1}{c}{}  & 1 & CaF$_{2}$-1  &  $125 - 179$ & 44.1  & 150.94        \\
\multicolumn{1}{c}{}  & 2 & BaF$_{2}$  & $133 - 183$ & 37.8  &  154.96  \\
\multicolumn{1}{c}{}  & 3 & Sapphire  & $145 - 181$ & 27.4&  160.7   \\
\multicolumn{1}{c}{}  & 5 & Silica    &  $160 - 179$ & 13.13&  170.3    \\
\multicolumn{1}{c}{}  & 7 & CaF$_{2}$-2  &  $126 - 179$ & 42&  151.7    \\ 
\hline
\multirow{6}{*}{NUV}  & 0 & Block     &        &      \\
& 1 & Silica           & $194 - 304$ &  76.9   &   241.8          \\
& 2 & NUV15           &  $190 - 240$ & 27.1       & 218.5     \\
& 3 & NUV13           &  $220 - 265$ &  28.17     &    243.6          \\
& 5 & NUVB4           &  $245 - 282$ &  28.23      &   262.8     \\
& 6 & NUVN2           &  $273 - 288$ & 8.95       &   279.0     \\
\hline
\multirow{6}{*}{VIS}    & 0 & Block        &     &   &     \\
& 1 & VIS3     & $385 - 530$    & 107.65  & 458.6     \\
& 2 & VIS2 & $360 - 410$   &  36.24   &  390.5   \\
& 3 & VIS1     & $318 - 374$  & 37.65  & 347.4    \\
& 4 & Neutral Density & $366 - 533$ &   97.0  &  450.1  \\
& 5 & BK7 Window      & $304 - 550$ & 185.   &  430.7     \\ 
\hline
\end{tabular}
\end{table*}

UVIT consists of two identical Ritchey-Chr\'etien telescopes with intensified CMOS detectors: one telescope with a CsI detector (FUV channel) with the other telescope feeding two detectors through a dichroic. The dichroic reflects light in the NUV spectral range onto a CsTe photocathode and transmits the visible light onto an S20 photocathode. Each channel is equipped with a filter wheel with 5 filters providing spectral coverage in a number of passbands from the FUV to the visible (Table~\ref{filterprop}). The effective bandwidth in Table~\ref{filterprop} is the integral of the normalized effective area, with the central (or `mean') source-independent wavelength defined as
\begin{equation}
\lambda_0=\frac{\int \lambda A_{\rm norm}(\lambda)d\lambda}{\int A_{\rm norm}(\lambda)d\lambda}\,,
\label{central_lambda}
\end{equation}
where $A_{\rm norm}$ is the effective area normalized to 1. We have plotted the effective areas measured during the ground calibrations (available at {\tt http://uvit.iiap.res.in/Instrument/Filters}) in Fig.~\ref{fig:spectra}. 

\section{Calibration}

\subsection{Overview}

The UVIT instrument was calibrated in the M. G. K. Menon Space Science Laboratory at the Indian Institute of Astrophysics with results described by \cite{Kumar2012}. The first 6 months of the mission were dedicated to the performance and verification (PV) phase when selected targets were viewed through different filter combinations. These observations and their analysis and reduction have been discussed by \cite{Subramaniam2016} and \citet{Tandon2017a}. We have used these observations along with our own Guaranteed Time (GT) and AO observations to independently characterize the in-flight performance of the UVIT instrument. All our observations begin with the Level 1 data created by the Indian Space Science Data Centre (ISSDC). We have processed these data through the {\it JUDE} software \citep{Murthy_jude2016, murthy2017} to create photon event lists and images of the sky.

\subsection{Photometric Calibration}

\subsubsection{Standard Calibrators}

Two standard stars were observed as part of the UVIT calibration program: white dwarfs LB227 and HZ4, both of which are standard calibrators of the instruments on the {\it Hubble Space Telescope} \citep{bohlinHST}. We obtained their spectra from the CALSPEC database\footnote{http://www.stsci.edu/hst/observatory/crds/calspec.html} and smoothed with a natural cubic spline. The spectra of the two stars are plotted in Fig.~\ref{fig:spectra} along with the effective area curves of the UVIT FUV and NUV filters and both \galex bands. The UVIT effective areas are from {\tt http://uvit.iiap.res.in/Instrument/Filters} and the \galex effective areas from the SVO Filter Profile Service \citep{Rodrigo2012}. We convolved the stellar spectrum with the effective area curve of each filter and have tabulated the expected count rate in Table~\ref{table:calibstars}. 

The standard operating procedure for the FUV and NUV detectors is to read the $512\times 512$ full frame at a rate of 29 frames per second; faster readouts (up to 600 per second) with smaller windows are available but we only used full-frame observations here. For the purposes of this work, we have defined an observation as a contiguous set of data frames; i.e. we did not co-add different data sets where, for whatever reason, there was a time gap of more than 10 seconds in the data. With this definition, we have 59 independent observations of HZ4 over 4 orbits, and 22 observations of LB227 in one orbit taken over the period Feb. to Dec. 2016. We measured the count rate in each of the observations and tabulated the weighted mean and standard deviation in Table~\ref{table:calibstars}, assuming Poissonian statistics in the observed counts. We compared the observed count rate to that expected from the ground calibration and found that it was between 70\% -- 90\% of the pre-flight values (column `Obs./Exp.' in Table~\ref{table:calibstars}). Note that the HZ4 was too bright to observe in most of the NUV filters.

\begin{table*}
\centering
\caption{UVIT count rates (counts s$^{-1}$) for standard stars.}
\label{table:calibstars}
\begin{tabular}{|l|l|l|l|c|c|c|c|c|c|c|c|c|}
\hline
\multicolumn{3}{|c|}{\multirow{2}{*}{Filter}} & \multicolumn{2}{c|}{HZ4 expected }& \multicolumn{2}{c|}{HZ4 observed} & \multicolumn{1}{c|}{\multirow{2}{*}{Obs./Exp.$^{c}$}} & \multicolumn{2}{c|}{LB227 expected} & \multicolumn{2}{c|}{LB227 observed} & \multicolumn{1}{c|}{\multirow{2}{*}{Obs./Exp.$^{c}$}} \\ 
\cline{4-7} \cline{9-12}
\multicolumn{3}{|c|}{} & \multicolumn{1}{c|}{} & \multicolumn{1}{c|}{Corr.$^{a}$} & \multicolumn{1}{c|}{mean} &  \multicolumn{1}{c|}{stdev} & \multicolumn{1}{c|}{} & & \multicolumn{1}{c|}{Corr.$^{a}$} & \multicolumn{1}{c|}{mean}   & stdev &\multicolumn{1}{c|}{}\\ \hline
\multicolumn{1}{|c|}{\multirow{5}{*}{FUV}} & CaF$_2$-1 & F1 & 28.94 & 20.59 & 15.92 & 0.19 & 0.77 & 15.69 & 14.17 & 10.96 & 0.41 & 0.77 \\ 
 & BaF$_2$ & F2 & 26.34 & 19.6 & $14.75^d$ & - & 0.75 & 13.74 & - & 9.85 & 0.08 & 0.72 \\ 
 & Sapphire & F3 & 17.61 & 15.34 & 12.13 & 0.22 & 0.79 & 8.95 & - & 7.60 & 0.06 & 0.85 \\ 
 & Silica & F5 & 6.33 & - & 5.12 & 0.13 & 0.81 & 3.09 & - & 2.99 & 0.06 & 0.97 \\ 
& CaF$_2$-2 & F6 & 25.88 & 19.41 & - & - & - & 13.97 & - & - & - & - \\ 
\midrule
\multicolumn{1}{|c|}{\multirow{5}{*}{NUV}} & Silica & F1 & 133.55   & - &  $b$ 
 & - & - & 68.75 & - & - & - & - \\ 
& B15 & F2 & 9.92 & - & 6.88 & 0.49 & 0.69 & 5.13 & - & 3.54 & 0.07 & 0.69 \\ 
& B13 & F3 & 49.77 & - &  ${b}$  & - & - & 25.65 & 19.32 & 14.34 & 0.19 & 0.74 \\ 
& B4 & F5 & 35.59 & - &  ${b}$  & - & - & 18.62 & 15.92 & 11.31 & 0.07 & 0.71 \\  
& N2 & F6 & 6.02 & - & 4.84 & 0.27 & 0.80 & 3.13 & - & 2.55 & 0.04 & 0.82 \\ 
\hline
\multicolumn{12}{l}{$^{a}$Non-linearity correction applied (see Sec.~\ref{sec:nonlinearity}).}\\
\multicolumn{12}{l}{$^{b}$ Window mode.}\\
\multicolumn{12}{l}{$^{c}$ Ratio between observed and expected count rates.}\\
\multicolumn{12}{l}{$^{d}$ Single observation.}\\
\end{tabular}
\end{table*}

\subsubsection{Calibration using \galex data}

In principle, the photometric calibration should be done by comparing the observed fluxes to those predicted from the standard stars. However, because only two standard stars were observed in the PV phase (HZ4 and LB227), we have expanded the list to include those stars detected in the FUV and the NUV which were also detected by \galex. The \galex photometric calibration has been described by \cite{Morrissey2007} and was tied to an absolute photometric calibration through observations of hot white dwarfs, including HZ4 and LB227.

\begin{figure}
\centering
\includegraphics[width=4cm]{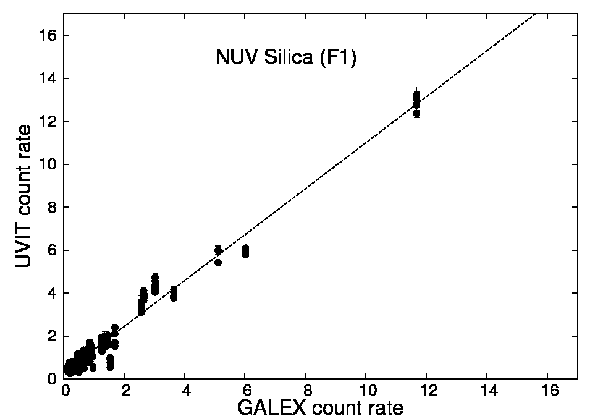}
\includegraphics[width=4cm]{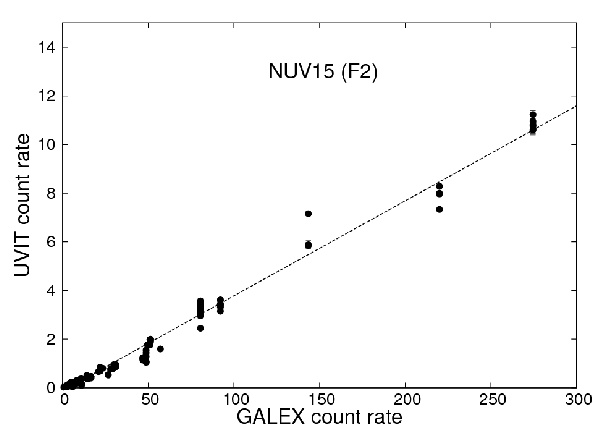}
\includegraphics[width=4cm]{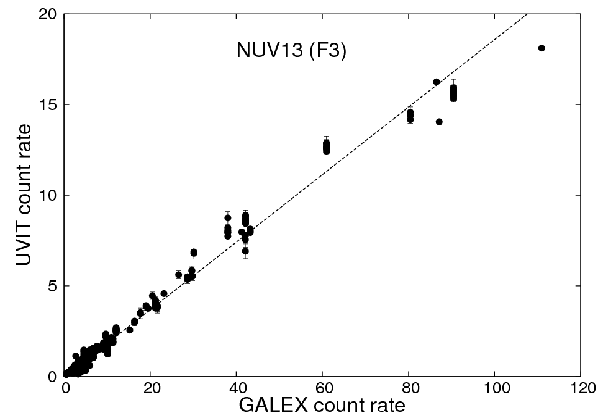}
\includegraphics[width=4cm]{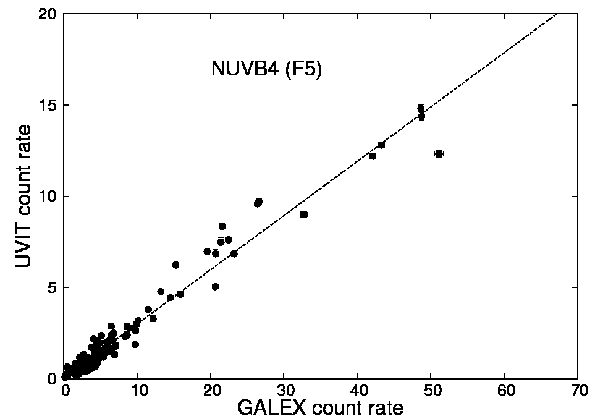}
\includegraphics[width=4cm]{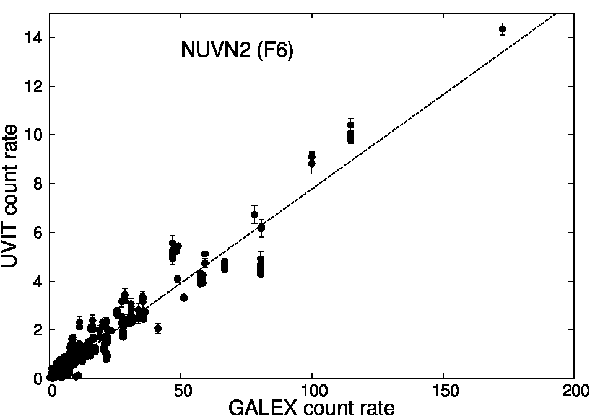}
\caption{Comparison of UVIT photometry with \galex in different UVIT NUV filters.}
\label{fig:nuv}
\end{figure}

\begin{figure}
\centering
\includegraphics[width=4cm]{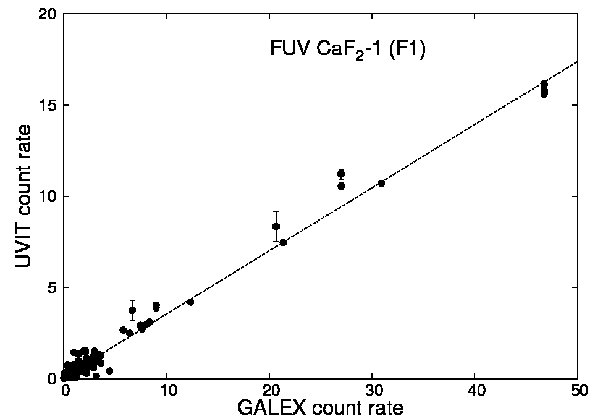}
\includegraphics[width=4cm]{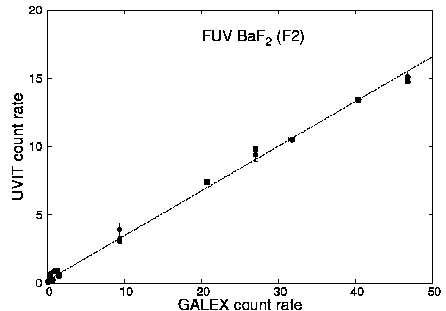}
\includegraphics[width=4cm]{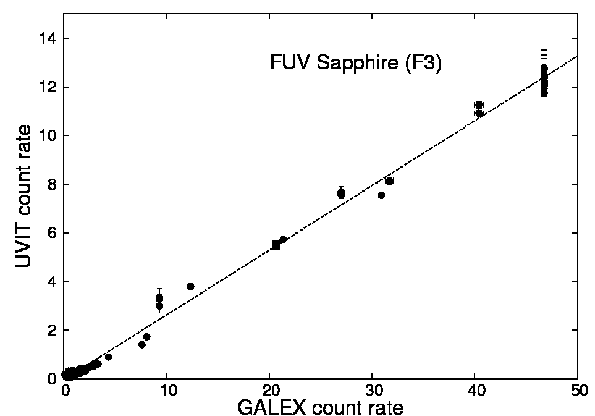}
\includegraphics[width=4cm]{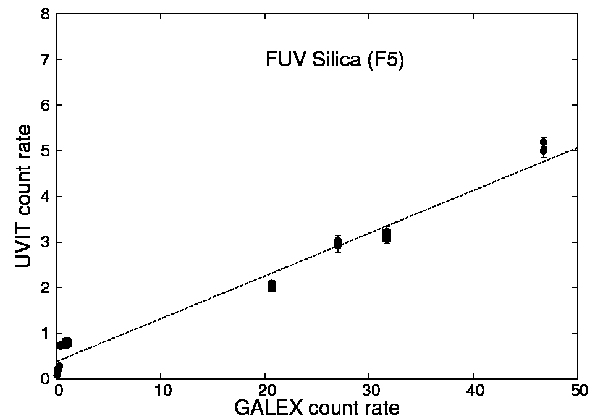}
\caption{Comparison of UVIT photometry with \galex in different UVIT FUV filters.}
\label{fig:fuv}
\end{figure}

We used the IDL (Interactive Data Language\footnote{http://www.harrisgeospatial.com/ProductsandTechnology/Software/IDL.aspx}) library routine {\it find.pro} (adapted from DAOPHOT: \citet{Stetson1987}) to find the point sources in the UVIT images and matched them with \galex\ns. We then used {\it aper.pro} to perform aperture photometry and extract fluxes from the UVIT images and both \galex bands, where available. If there were multiple \galex observations of a field, we used the one with the greatest exposure time. In each case, we inspected the image to ensure that we were selecting the same source in both UVIT and \galex\ns. Stars used in our calibration are listed in Table~\ref{table:phottable}.

\begin{table*}
\caption{Photometry of stars used in the calibration. The full table is available in the electronic attachment.}
\label{table:phottable}
\begin{tabular}{lllccccll}
\hline
Star ID & \multicolumn{1}{c}{\begin{tabular}[c]{@{}c@{}}RA\\  {[}deg{]}\end{tabular}} & \multicolumn{1}{c}{\begin{tabular}[c]{@{}c@{}}Dec\\ {[}deg{]}\end{tabular}} & \multicolumn{1}{c}{\begin{tabular}[c]{@{}c@{}}\galex\\ {[}counts s$^{-1}${]}\end{tabular}} & \multicolumn{1}{c}{\begin{tabular}[c]{@{}c@{}}$t_{\galex\ns}$\\ {[}sec{]}\end{tabular}} & \begin{tabular}[c]{@{}c@{}}UVIT\\ {[}counts s$^{-1}${]}\end{tabular} & \multicolumn{1}{c}{\begin{tabular}[c]{@{}c@{}}$t_{\rm UVIT}$\\ {[}sec{]}\end{tabular}} & Fliter & Detector \\ \midrule
1 & 11.9652 & 85.3188 & 20.664 & 222 & 8.347 & 12.801 & F1 & FUV \\
2 & 62.3706 & 17.1315 & 26.998 & 14807.95 & 9.89 & 326.072 & F2 & FUV \\
3 & 58.842 & 9.7884 & 46.752 & 13321.1 & 15.039 & 71.248 & F2 & FUV \\
4 & 20.9410 & -58.8057 & 40.404 & 219 & 11.257 & 630.811 & F3 & FUV \\
5 & 62.3706 & 17.1315 & 26.998 & 14807.95 & 2.968 & 244.033 & F5 & FUV \\
\midrule
1 & 10.6417 & -9.2020 & 11.681 & 28993.15 & 12.368 & 297.733 & F1 & NUV \\
2 & 12.0834 & 85.2239 & 92.103 & 222 & 3.149 & 255.785 & F2 & NUV \\
3 & 83.5602 & 21.9034 & 21.218 & 167 & 3.997 & 548.825 & F3 & NUV \\
4 & 12.951 & -27.1692 & 43.264 & 224 & 12.813 & 1603 & F5 & NUV \\
5 & 256.536 & 78.624 & 80.734 & 1675.05 & 6.170 & 48.7494 & F6 & NUV \\ \hline
\end{tabular}
\end{table*}

There is a tight correlation between the observed UVIT counts (in all bands) and the \galex counts up to an observed count rate of 15 cps in UVIT above which non-linearity sets in (about 9.7\% roll-off, as discussed below). There is effectively no non-linearity in the \galex data at these fluxes because of the faster response time of their delay-line anodes. The errors in either data set are dominated by photon noise and were calculated from the square root of the total number of counts. We have used the IDL routine {\it fitexy.pro} which handles errors in both $x$ and $y$ to calculate the slope and the uncertainty between the UVIT and \galex fluxes in each filter. These are tabulated in Table~\ref{table:slope} and plotted in Figs.~\ref{fig:nuv} and \ref{fig:fuv}. The two broadband FUV filters (F1: CaF$_{2}$ and F2: BaF$_{2}$) have a coverage similar to the \galex FUV band with an effective response of about 35\% of the \galex FUV response, as expected from the smaller (35-cm) UVIT primary mirror compared to the \galex primary (50~cm). The smaller UVIT mirror is compensated by the loss in responsivity in the \galex dichroic, and the response of the broadband NUV filter (F1: Silica) is close to that of the \galex NUV band.

We have made no assumptions about the spectral type of each star. This is unimportant for the broad-band filters where the filter response curve for both \galex and UVIT are similar but will impact the narrow-band filters where the source might have emission/absorption lines or the filters may have long tails leading to leakage from out of band counts. This is reflected in the scatter seen particularly in the  NUVN2 where the count rate is only about 7\% that expected in the \galex NUV band. 

We have converted the UVIT-\galex slopes into an absolute calibration using the \galex\ conversion factors of $1.40 \times 10^{-15}$ and $2.06 \times 10^{-16}$ erg cm$^{-2}$ s$^{-1}$ \AA$^{-1}$ (cps)$^{-1}$ in the FUV and NUV, respectively. The \galex calibration assumed the sources to be spectrally flat, regardless of the actual spectral type. This is obviously an approximation, and a correct calibration should include the spectral type of the source \citep{Ravichandran}. 
The slope was used as a scale factor to calculate the predicted UVIT count rate to estimate the effect of non-linearity. The scale factors are tabulated in Table~\ref{table:slope} and include the effects of the smaller bandpass of the narrow-band filters. Note that it is important to consider the spectral shape of the source when calculating the flux, particularly for the narrow-band filters. The flux $F(\lambda)$ can be derived from the counts using the following equation,
\begin{equation}
F(\lambda) = C \times CPS\,,
\label{eq:Fluxconv}
\end{equation}
where the conversion factors $C$ for UVIT for each filter are given in Table~\ref{table:slope}. These conversion factors (in units of erg cm$^{-2}$ \AA$^{-1}$ {\rm cnt}$^{-1}$) were derived using the UVIT/\galex slopes (Figs.~2 and 3) as follows
\begin{equation}
C_{\small \rm UVIT}=\frac{C_{\small \galex\ns}}{\rm slope}\,.
\label{eq:FluxconvUVIT}
\end{equation}

\subsubsection{Non-linearity}
\label{sec:nonlinearity}

\begin{figure}
\includegraphics[width=8cm]{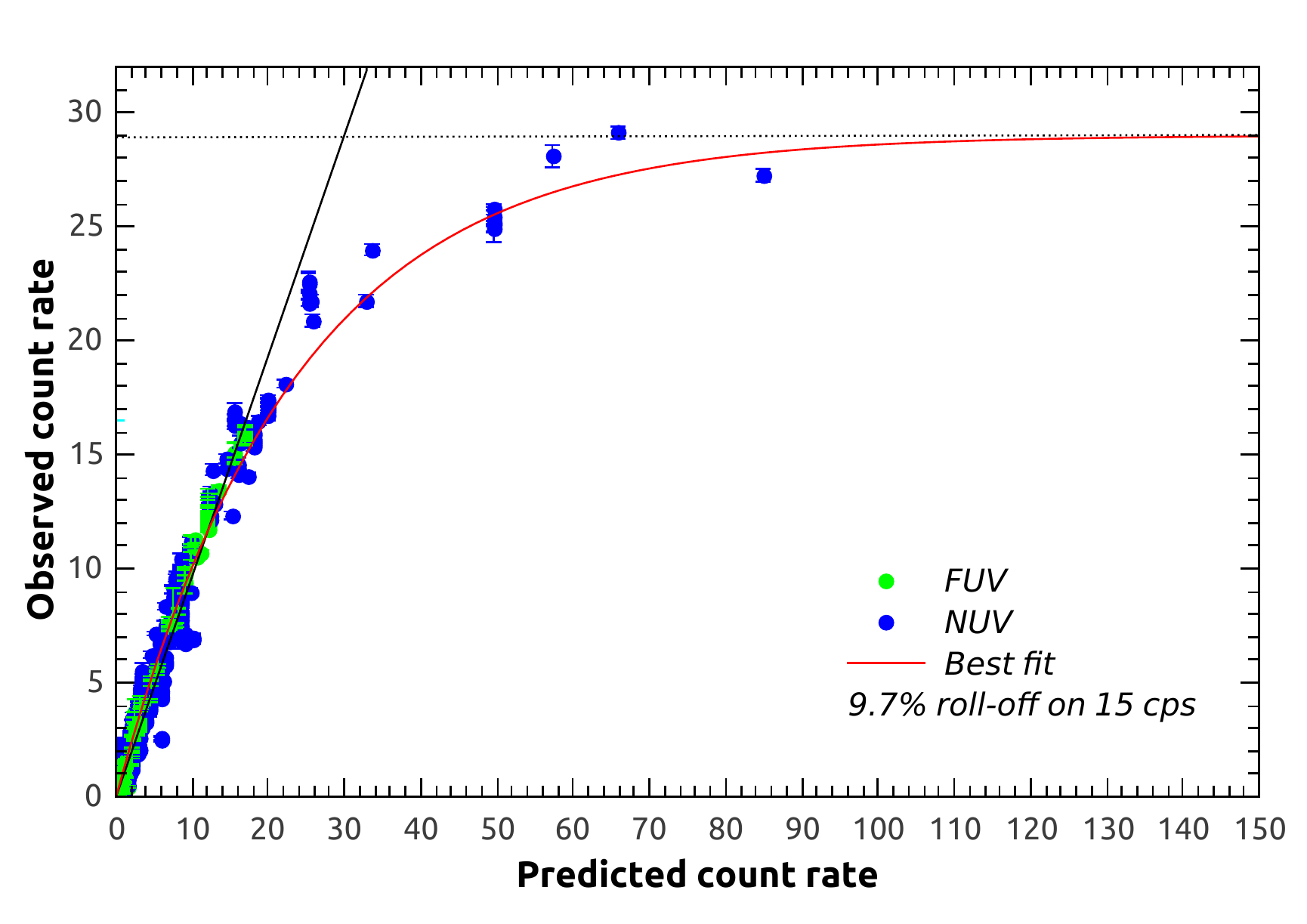}
\caption{Observed counts as a function of predicted counts for FUV (green points) and NUV (blue points).}
\label{fig:non-linearity}
\end{figure}

Intensified detectors are subject to non-linearity at high count rates because the detectors can only register one count per pixel per frame \citep{Fordham2000}. We have compared the observed UVIT count rates in both FUV and NUV channels for all filters (Fig.~\ref{fig:non-linearity}) with the scaled count rates (\galex count rate multiplied with slope in Table~\ref{table:slope}). We have used the formulation of \citet{Kuin2008} to model the non-linearity:
\begin{equation}
\text{C}_{\rm obs} = 29\times\left(1- \mathrm{e}^{-\alpha C_{\rm inc}/29}\right)\,,
\label{eq:non-linearity}
\end{equation}
where $\alpha = 1.24$ (determined empirically), $C_{\rm inc}$ is the number of events incident on the detector, $C_{\rm obs}$ is the number of events detected, and there are 29 frames in a second. Non-linearity in the observed counts sets in at 15 cps with about 9.7\% loss and may be corrected for up to 29 cps, at which the measured count rates saturate and the true counts can no longer be recovered. 

\begin{table*}
\caption{UVIT conversion factors (in erg cm$^{-2}$ \AA$^{-1}$ {\rm cnt}$^{-1}$).}
\label{table:slope}
\begin{tabular}{|l|l|l|c|c|c|c|c|}
\hline
Filter & Slot & Slope & Slope error & R      & Conversion factor & \citet{Tandon2017b} & Ratio \\
& &  \multicolumn{2}{c|}{  (UVIT/GALEX)} & &  & \\\hline
FUV CaF2\_1        & F1     & 0.3619 & 0.0013     & 0.9845 & 3.8689e-15  & 3.127E-15 & 0.81\\ 
FUV BaF2           & F2     & 0.3330 & 0.0018     & 0.9978& 4.2036e-15 & 3.593E-15 & 0.85\\ 
FUV Sapphire       & F3     & 0.2574 & 0.0008     & 0.9986 & 5.4399e-15 & 4.402E-15 & 0.81\\ 
FUV Silica        & F5     & 0.0980 & 0.0011     & 0.9848 & 1.4273e-14  & 1.071E-14 & 0.75\\ 
\midrule
NUV Silica         & F1     & 1.0586 & 0.0027    & 0.9873 & 1.9459e-16 & 2.270E-16 & 1.2\\ 
NUV B15            & F2     & 0.0353 & 0.0001     & 0.9956 & 5.8360e-15 & 5.356E-15 & 0.91   \\ 
NUV B13            & F3     & 0.1995 & 0.0005    & 0.9941 & 1.0327e-15 & 7.412E-16 & 0.71\\ 
NUV B4             & F5     & 0.2959 & 0.0014     & 0.9825 & 6.9611e-16 & 8.632E-16 & 1.24  \\ 
NUV N2             & F6     & 0.0736 & 0.0002     & 0.9723 & 2.7988e-15 & 3.577E-15  & 1.28\\ \hline
\end{tabular}
\end{table*}

\subsection{Temporal variation in sensitivity}

\begin{figure}
\centering
\includegraphics[width=8cm]{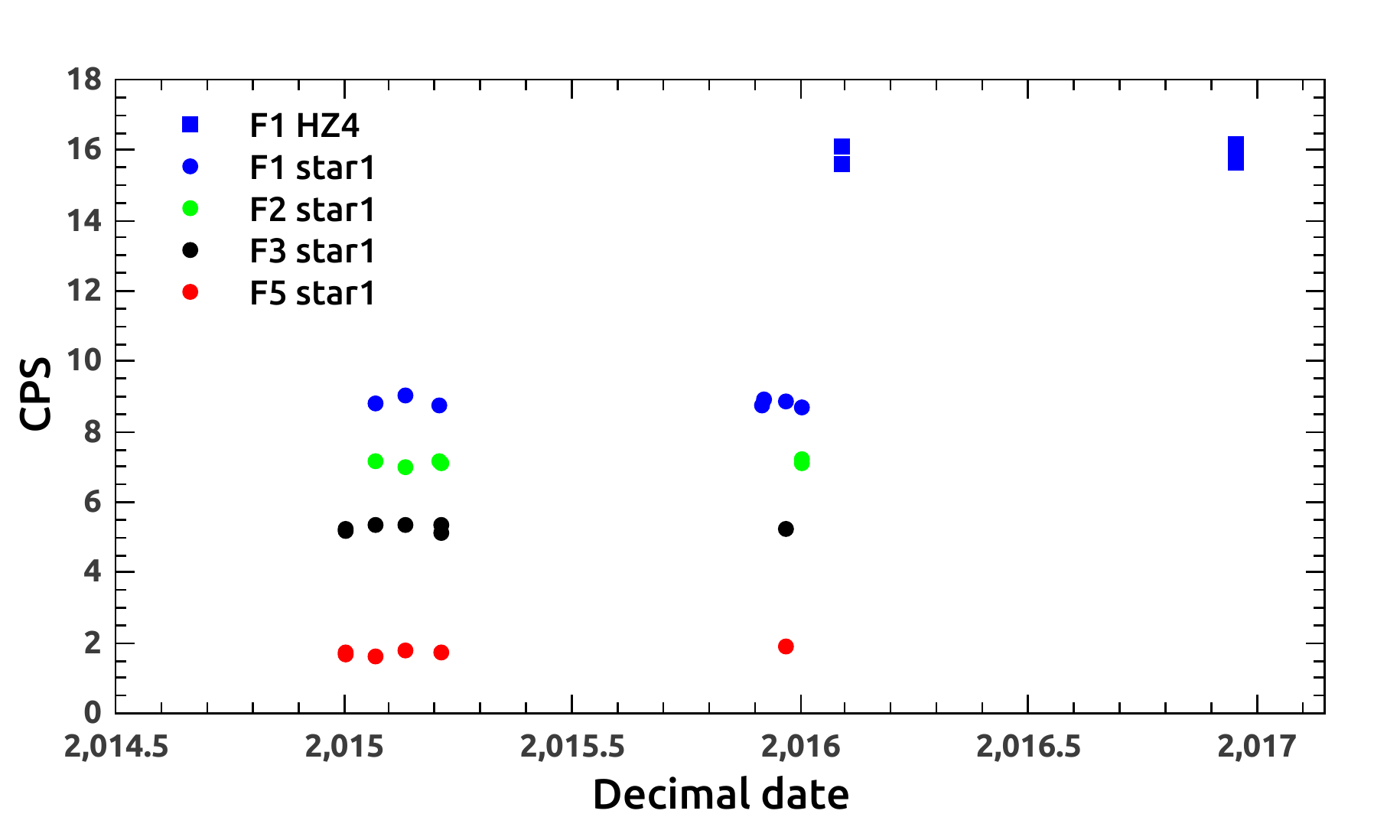}
\includegraphics[width=8cm]{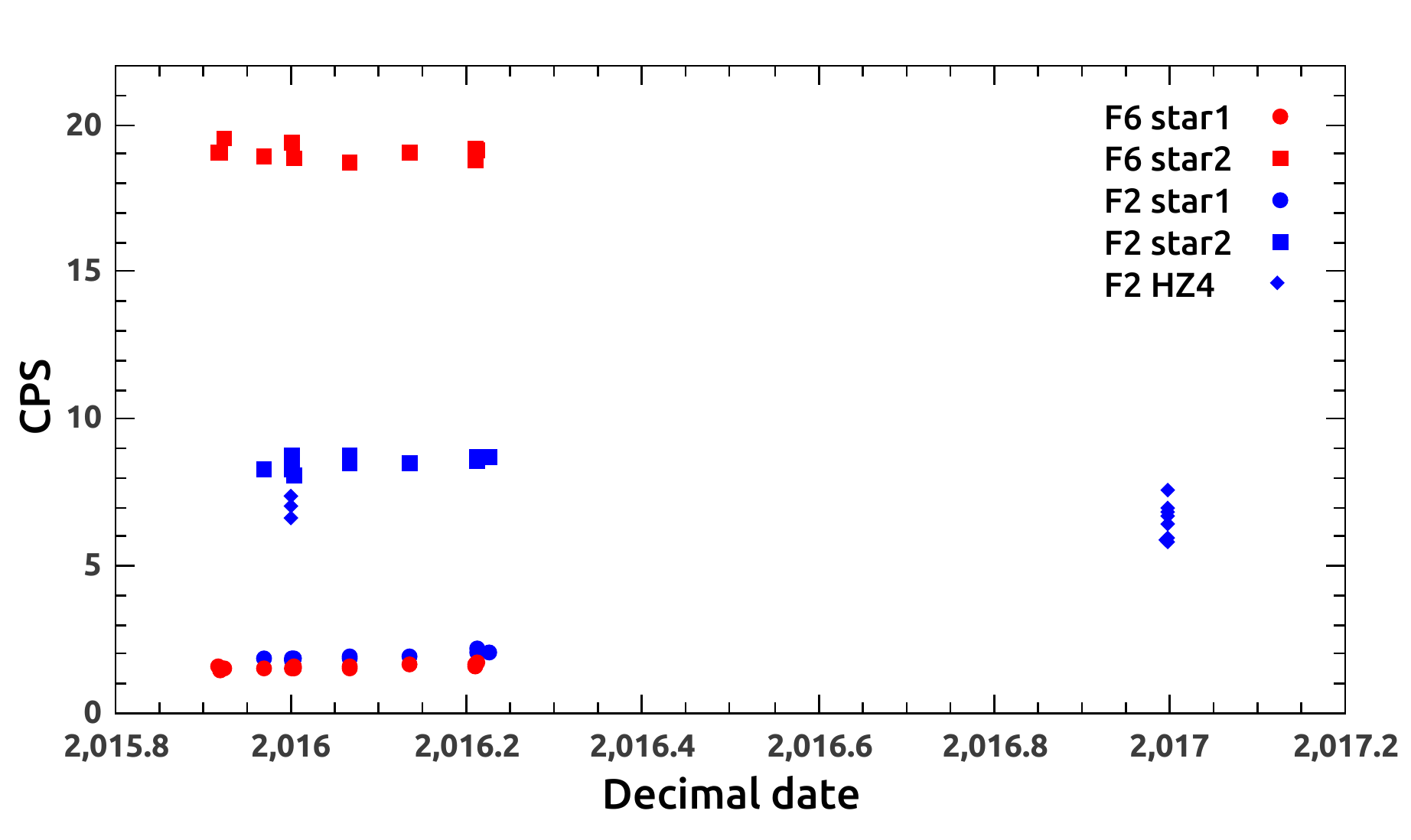}
\caption{UVIT count rate as a function of date. {\it Top}: CPS as a function of date of two stars (HZ4 and star 1) in FUV filters. {\it Bottom}: variation of CPS of HZ4 and star 2 in NUV filters.}
\label{fig:sensitivity_with_time}
\end{figure}

To estimate the possible loss in sensitivity over time, we have compared the counts of stars whose observations were performed over long enough baseline; in three FUV filters and in two NUV filters. We have used two bright stars in NGC 188 cluster (Star 1: BD+8412, star 2: NGC188~2091), in addition to HZ4, and plotted their count rates in Fig.~\ref{fig:sensitivity_with_time}. We find no evidence that the sensitivity has changed with time.

\subsection{Geometric Distortion} 

\begin{figure}
\centering
\includegraphics[width=8cm]{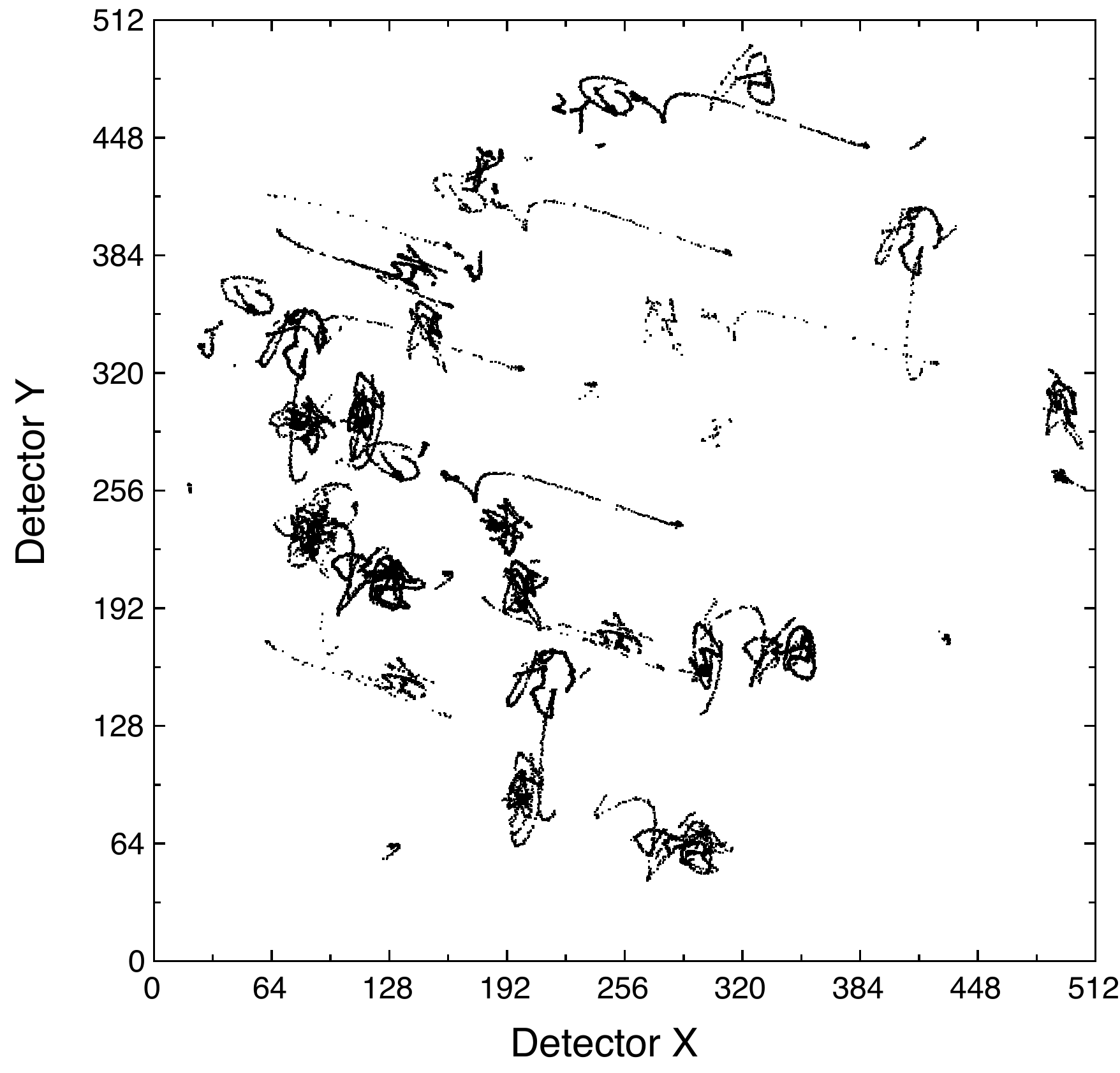}
\caption{Tracks of individual stars in the detector plane due to spacecraft motion over 9 orbits of NGC 188 in the NUV.}
\label{fig:distortion_x_y_pos}
\end{figure}

\begin{figure}
\includegraphics[width=8 cm]{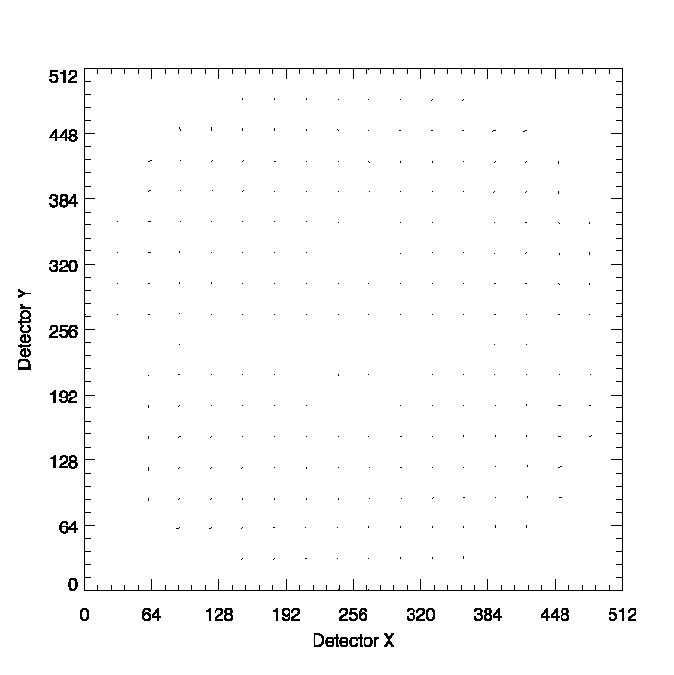}
\caption{Geometric distortion over the field of view of the NUV detector (full frame).}
\label{fig:dist_nuv}
\end{figure}

\begin{figure}
\centering
\includegraphics[width=8 cm]{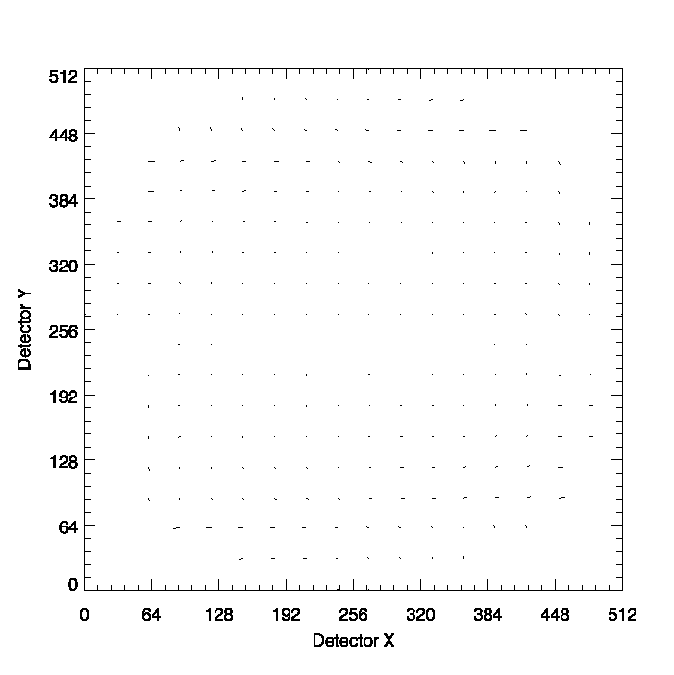}
\caption{Geometric distortion over the field of view of the FUV detector (full frame).}
\label{fig:dist_fuv}
\end{figure}

A ground measurement of the geometric distortion is available for the UVIT detectors alone, carried out before integration with the optical assembly \citep{girish2017}. The measurement was done using a grid of pinholes, the geometric configuration (i.e. pinhole spacing etc.) of which was calibrated using the IUCAA Faint Object Spectrograph \& Camera (IFOSC). The geometric distortion of IFOSC itself is only known via simulations. \citet{girish2017} found a complex distortion pattern and reported the improvement of the astrometry in VIS flight images after applying their distortion correction. This is true, in principle, as one of the limiting factors for the achievable resolution is the spacecraft motion, which is corrected for by applying a shift in $x$ and $y$. This shift will be affected by the geometric distortion and hence the resolution may not be uniform over the entire detector plane. 

Ideally, geometric distortion would have been corrected in-flight through observations of open clusters such as NGC 188 but spacecraft motion made it impossible to correlate the positions on the detector plane with the distortion. Instead, we selected three relatively bright stars in the FUV and nine in the NUV observations of NGC 188 and calculated their centroids at intervals of one second. The individual star trails are plotted in Fig.~\ref{fig:distortion_x_y_pos} for all NUV observations of NGC 188.

We used the standard SIP (Simple Imaging Polynomial) formulation \citep{Shupe2005}
\begin{align}
u = x + &A_{20}(x - 256)^2 + A_{11}(x-256)(y-256) \nonumber\\
+ &A_{02}(y-256)^2\,, \nonumber \\
v = y + &B_{20}(y - 256)^2 + B_{11}(x-256)(y-256) \nonumber\\
+ &B_{02}(x-256)^2\,.
\label{eq:distcoef}
\end{align}
to correct the ($x, y$) pairs in detector coordinates with the centre at ($0,0$) into the corrected $u-v$ plane. The angles between stars will remain constant in the undistorted plane, regardless of spacecraft motion, and we determined the coefficients of distortion by forcing the distance between stars in the $u-v$ plane to be the actual angular distance. There is considerable noise in calculating the distortion because of the rapidity of the spacecraft motion and the intrinsic photon noise of the observations in the short time per pixel but we have found a good convergence in the derived coefficients of distortion (Table~\ref{table:distcoef}). The distortion maps derived from Eq.~\ref{eq:distcoef} for the NUV and FUV detectors are shown in Figs.~\ref{fig:dist_nuv} and ~\ref{fig:dist_fuv}, respectively.

\begin{table}
\centering
\caption{Distortion coefficients from Eq.~\ref{eq:distcoef} for NUV and FUV channels.}
\label{table:distcoef}
\begin{tabular}{|c|c|c|}
\hline
Coefficient &NUV        & FUV        \\ 
\hline
$A_{20}$&-3.7e-05 & -4.3e-05 \\ \hline
$A_{11}$&-4.4e-05 & -7.1e-05 \\ \hline
$A_{02}$&1.8e-05  & 1.0e-04     \\ \hline
$B_{20}$&-2.7e-05 & -3.4e-05 \\ \hline
$B_{11}$&-6.2e-05 & -5.9e-05 \\ \hline
$ B_{02}$&2.2e-05  & 2.8e-05  \\ \hline
\end{tabular}
\end{table}

Although \citet{girish2017} indicate that applying a distortion correction to the data improves the resolution of the instrument, we find that the effect is small with no measurable improvement in the instrument PSF. We recommend co-adding the frames in an observation and performing a distortion correction as part of the astrometric solution where the signal-to-noise ratio is better.

\subsection{Flat Fielding}

\begin{figure}
\centering
\includegraphics[width=8cm]{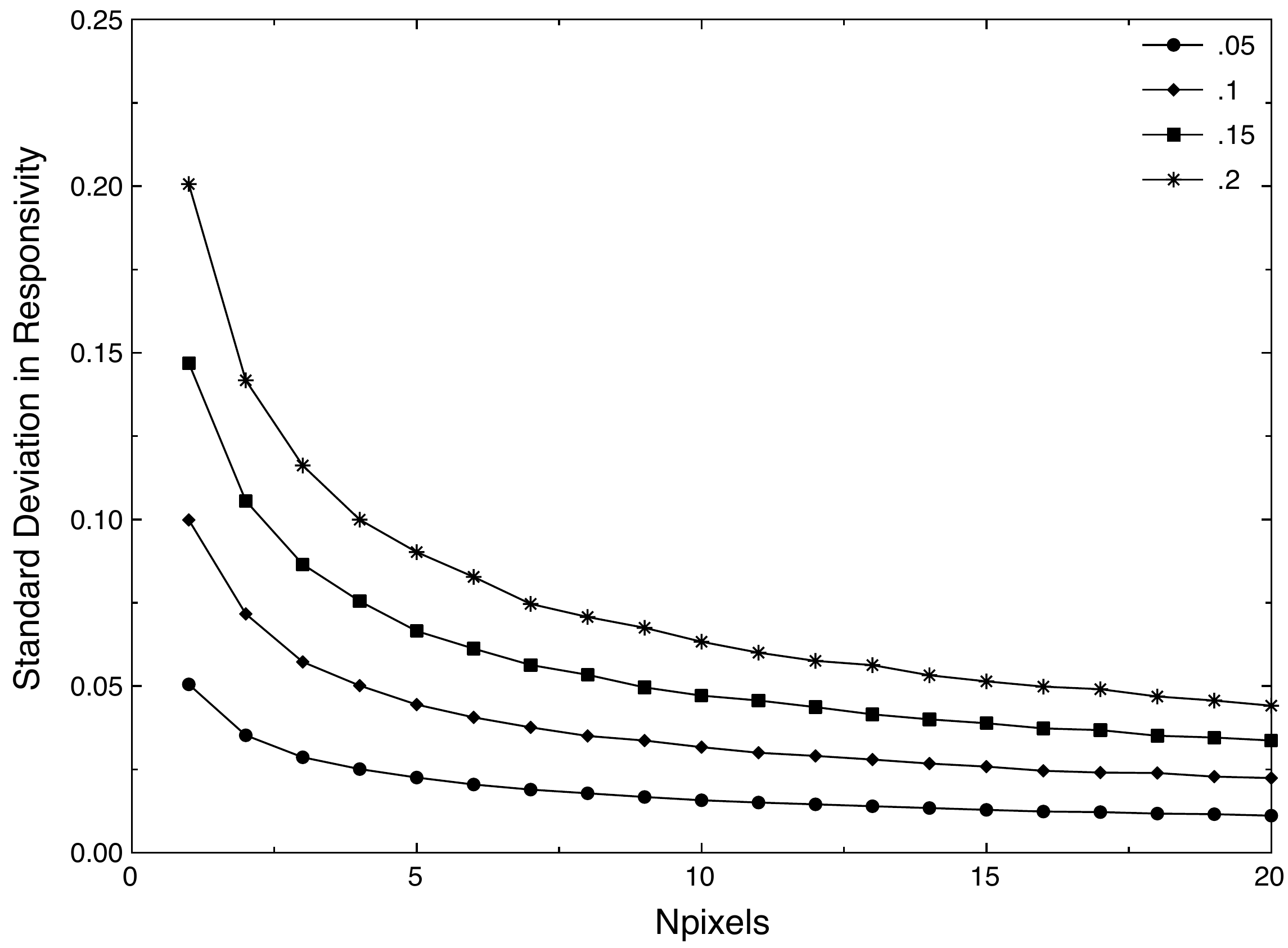}
\caption{Expected variation in stellar flux due to random fluctuations in pixel sensitivity.}
\label{fig:flat_sens}
\end{figure}

\begin{figure}
\centering
\includegraphics[width=8cm]{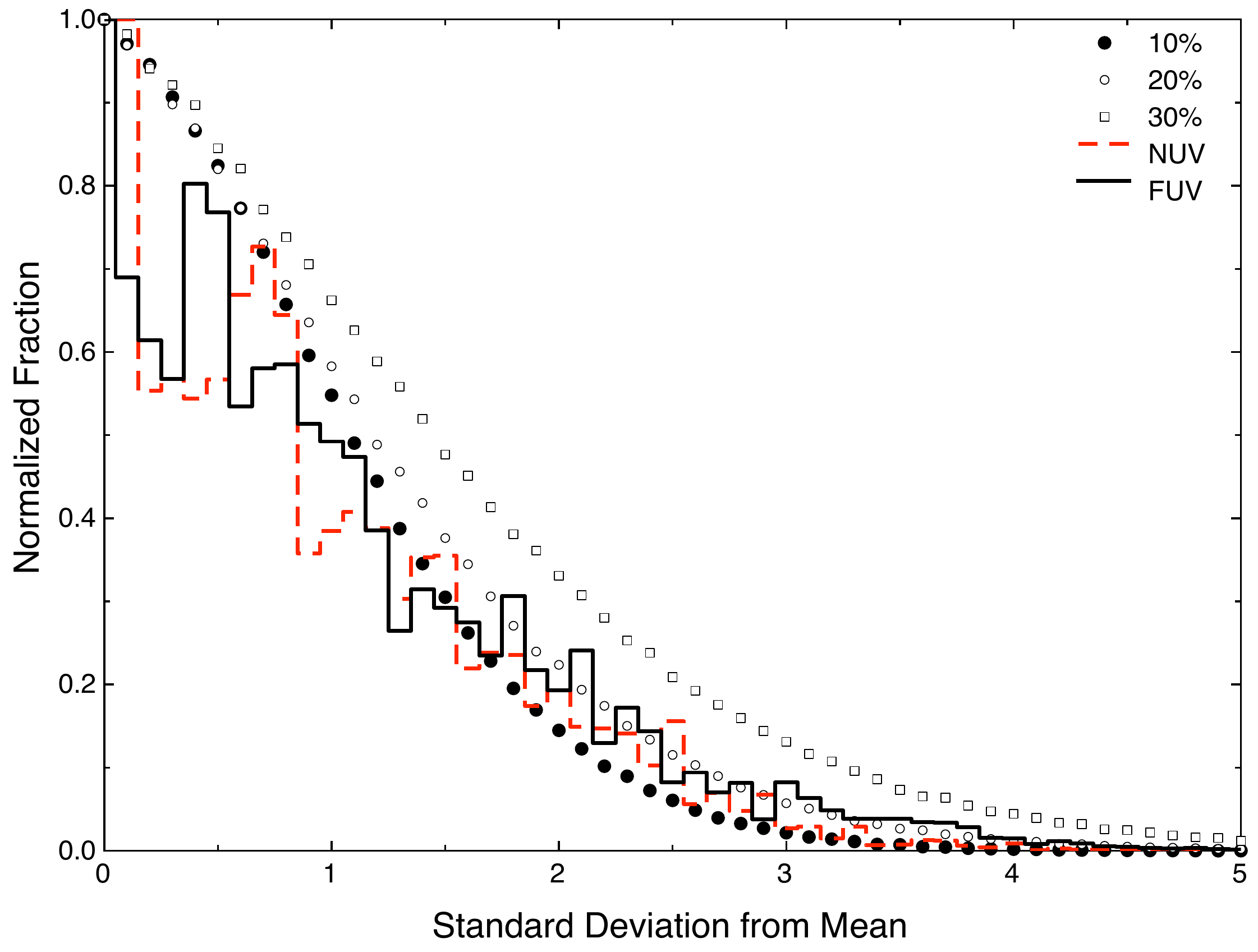}
\caption{Number of $\sigma$ away from the mean for the observations (histogram) and different levels of non-uniformity.}
\label{fig:flat_field_total}
\end{figure}

The flat field correction accounts for pixel to pixel variations in the sensitivity across the detector but are difficult to determine from in-flight observations because photon counting statistics dominate the signal. We will explore the flat-fielding below but will demonstrate that, in practice, corrections for the flat-field are unimportant in the context of UVIT because any source is averaged over many pixels due to spacecraft motion.

We have run a number (10,000) of simulations in which we assumed that the point sources were smeared over $N$ pixels, with the sensitivity in each pixel drawn from a normal distributions with a mean of 1 and $\sigma$ of 5 -- 20\%. The effective response that any star would see is the mean over the $N$ pixels over which it is smeared and, over a run of 10,000 simulations, will be close to 1. However, the standard deviation will translate into the uncertainty due to the non-uniform sensitivity for stars in different pixels. This is plotted in as a function of $N$ (the number of pixels) in Fig.~\ref{fig:flat_sens} and suggests that differences of 10\% in the sensitivity between pixels would result in an uncertainty of about 2\% if the star is smeared over 20 pixels, as is the case for most UVIT observations. We will discuss this in the context of actual observations below.

The most obvious measure of variations in the sensitivity of the detector is the observed count rate for a given source as it moves in the detector plane. We have already tracked the positions of 3 stars in the FUV and 9 in the NUV observations of NGC 188, and recorded the counts for each pixel in the context of our derivation of the geometric coefficients. These counts will vary because of photon noise and because of the non-uniformity, and we have plotted the deviations from the mean for each pixel in Fig.~\ref{fig:flat_field_total}. We then ran a number (10,000) of Monte Carlo simulations where the variation in the count included both photon noise and sensitivity variations of 10\%, 20\% and 30\% per pixel. At this stage in our data analysis, we can only say that the variations in the flat field are less than 10\% per pixel, and that it is not necessary to use a flat field in extracting fluxes because of the motion of the spacecraft.

\section{PSF}

The intrinsic point spread function of the UV detectors is expected to be 1.8\arcsec\ \citep{Kumar2012} but is affected by the spacecraft registration. The primary method of data registration is to use the VIS images in which there are more stars but these have a time resolution of 2.5 seconds during which the spacecraft may move over a significant number of pixels. We have developed a new method \citep{Murthy_jude2016, murthy2017} in which we follow the centroid of a star in the UV images, themselves, in which the time resolution is determined by the brightness of the star but can be as good as 0.35 seconds.

\begin{figure}
\centering
\includegraphics[scale=0.4]{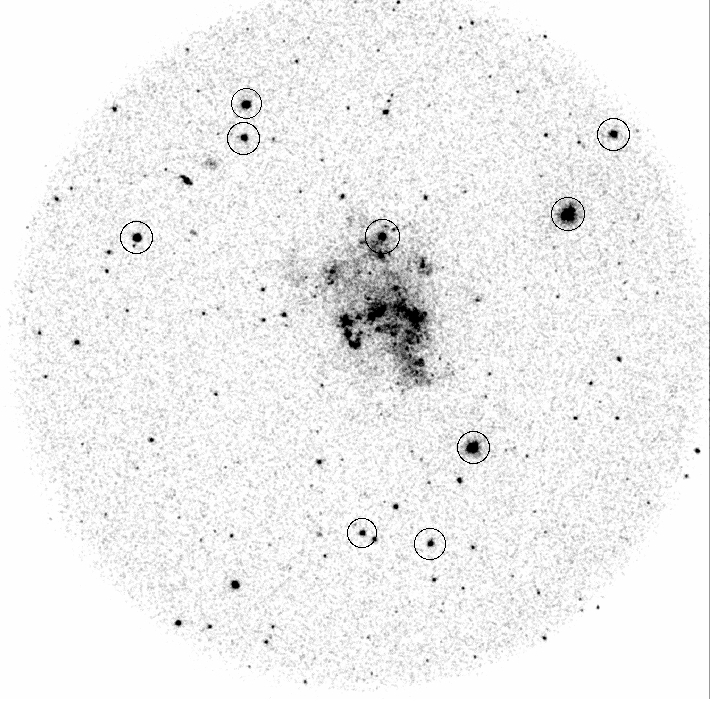}
\caption{NUV image of Holmberg-II taken on 30 Sept. 2016 with co-added exposure of 1194.6 sec. In black circles are the PSF stars.}
\label{fig:holmberg_psf}
\end{figure}

We have used PV observations of NGC 188 and GT observations of Holmberg-II galaxy (e.g. Fig.~\ref{fig:holmberg_psf}) in which there are a number of stars of different brightness and at least one bright star that we can use to correct for spacecraft motion. We used the {\it mpfit2dpeak} function in IDL to fit a 2-D Gaussian profile to each of the stellar profiles and calculated the FWHM in both $x$ and $y$ directions. The FWHM is 2.3 pixels (0.97\arcsec ) in the best case, but is more typically in the range from 3 -- 4 pixels (1.2 -- 1.6\arcsec). We have found no evidence for any spatial variation of the PSF over the detector plane. 

\section{Conclusions}

We have performed an independent evaluation of the performance of the UVIT FUV and NUV instruments based on their in-flight performance. We find that the performance is close to that expected from the ground-based calibration. The photometric sensitivity is about 35\% that of \galex in the FUV broad band filters and about the same as \galex in the NUV broad band filter. We find that the resolution can be as good as 1.2 -- 1.6\arcsec. Flat-fielding is unimportant for UVIT largely because the spacecraft moves enough during an observation that any variations are smeared out. We have derived a distortion correction but since the data are noisy, we recommend that the distortion correction be done as part of the astrometric correction post-processing.

\citet{Tandon2017b} have determined somewhat different calibration factors (Table~\ref{table:slope}) using only HZ4. This has the advantage that the spectrum of the star is known but is in the non-linear regime in most bands. Those bands with the highest count rates were observed with a high frame rate mode for which the timing was uncertain. We have chosen a broader selection of stars in the linear range of the detectors and tied our calibration to \galex calibration with the assumption that the individual stars will have the same relative response. This appears to be a good approximation given that we obtain excellent correlations between the count rates in both instruments (Figs.~\ref{fig:nuv} and~\ref{fig:fuv}) and we believe that, because of the brightness of HZ4, our values better represent the response of the instrument.

Our results serve as a validation of both the UVIT processing software and our alternative set of tools \citep{Murthy_jude2016}. UVIT is beginning to reach its potential and with the opening of the satellite to guest observers, including the international community, we may expect a flood of results in the near future. We will provide support to anyone who would like to use our software, or our results.

\section*{Acknowledgements}

This research has made use of the Spanish Virtual Observatory (SVO) Filter Profile Service (http://svo2.cab.inta-csic.es/theory/fps/) supported from the Spanish Ministry of Economy and Competitiveness (MINECO) through grant AyA2014-55216. We also acknowledge the Gnu Data Language (GDL), the Interactive Data Language ( IDL) Astronomy Library and its many contributors. This research has made use of National Aeronautics and Space Administration (NASA) Astrophysics Data System Bibliographic Services. Many people at Indian Institute of Astrophysics (IIA),  Indian Space Research Organisation (ISRO), Inter-University Centre for Astronomy and Astrophysics  (IUCAA), Tata Institute of Fundamental Research (TIFR), National Research Council (NRC,Canada) and University of Calgary have contributed to different parts of the spacecraft, instrument and the operations.

Some/all of the data presented in this paper were obtained from the Mikulski Archive for Space Telescopes (MAST). Space Telescope Science Institute (STScI) is operated by the Association of Universities for Research in Astronomy, Inc., under National Aeronautics and Space Administration (NASA) contract NAS5-26555. Support for MAST for non-Hubble Space Telescope (HST) data is provided by the NASA Office of Space Science via grant NNX09AF08G and by other grants and contracts.

This research has been supported by the Department of Science and Technology (DST) under grants no. SR/S2/HEP-050/2012 dated 14-08-2013 to Christ University and EMR/2016/00145 to  Indian Institute of Astrophysics (IIA).
\bibliographystyle{mnras}
\bibliography{uvit_cal}



\label{lastpage}
\end{document}